\documentclass[sigconf]{acmart}

\usepackage{booktabs} 
\usepackage{tabularx}
\usepackage{censor}

\setcopyright{rightsretained}

\acmDOI{}
\acmISBN{}

\acmConference[LBRS@RecSys '18]{the Late-Breaking Results track part of the Twelfth ACM Conference on Recommender Systems}{October 2018}{Vancouver, BC, Canada}
\acmYear{2018}
\copyrightyear{2018}

\begin{document}
\title{Predicting Musical Sophistication from Music Listening Behaviors: A Preliminary Study}

\author{Bruce Ferwerda}
\affiliation{%
  \institution{Department of Computer Science and Informatics}
  \institution{J\"{o}nk\"{o}ping University}
  \city{J\"{o}nk\"{o}ping}
  \state{Sweden}
}
\email{bruce.ferwerda@ju.se}

\author{Mark Graus}
\affiliation{%
  \institution{Department of Marketing and Supply Chain Management}
  \institution{Maastricht University}
  \city{Maastricht}
  \country{the Netherlands}
}
\email{mp.graus@maastrichtuniversity.nl}

\renewcommand{\shortauthors}{Ferwerda and Graus}

\begin{abstract}
Psychological models are increasingly being used to explain online behavioral traces. Aside from the commonly used personality traits as a general user model, more domain dependent models are gaining attention. The use of domain dependent psychological models allows for more fine-grained identification of behaviors and provide a deeper understanding behind the occurrence of those behaviors. Understanding behaviors based on psychological models can provide an advantage over data-driven approaches. For example, relying on psychological models allow for ways to personalize when data is scarce. In this preliminary work we look at the relation between users' musical sophistication and their online music listening behaviors and to what extent we can successfully predict musical sophistication. An analysis of data from a study with 61 participants shows that listening behaviors can successfully be used to infer users' musical sophistication. 
\end{abstract}


\maketitle

\section{Introduction \& Related Work}
There has been an increased interest in understanding online behaviors with psychological models and incorporating them to personalize systems (e.g.,~\cite{ferwerda2016personality}). Using psychological models to infer user characteristics from behavior has the advantage that personalization can be done without the need of additional, explicit data collection. Hence, it can be used to mitigate problems where data is scarce (e.g., the cold-start problem).

Most of the research on using psychological models in technological contexts rely on personality traits of users. Personality is a general model to categorize users and can even be used across domains~\cite{fernandez2015use}. However, using more domain dependent psychological models allows for more fine-grained identification of behaviors. Recent research are tapping into these domain dependent psychological models to improve personalizations. For example,~\citet{graus2018personalizing} showed that personalization based on parenting styles improved the overall user experience of an online parenting library. \citet{hauser2009website} exploited the cognitive styles of website users to provide a personalized experience. \citet{germanakos2016human} looked at learning styles to adapt the learning environment of students. 

In this preliminary work we look at the music domain and explore the influence of musical sophistication and the possibilities to infer musical sophistication of users from their music listening behavior. \citet{mullensiefen2014musicality} created a survey that measures musical sophistication which they define as "a psychometric construct that can refer to musical skills, expertise, achievements, and related behaviors across a range of facets that are measured on different subscales." This translates to that people with a higher degree of music sophistication in general engage more frequently in musical skills and behaviors, and have a greater and more varied repertoire of music behavior patterns. Hence, people's musical sophistication may well be reflected in their music listening patterns. In addition, musical sophistication has been suggested to be related to peoples' needs with regards to music systems~\cite{Celma2010}. The rise of online music services allows for tracking and analyzing music listening behavior on a larger scale. It provides opportunities to gain deeper insights on the relationships between music sophistication and listening behavior, as well as inferring musical sophistication from listening behaviors.

\section{Method}
The data for this study was collected as part of a larger study that investigated how the order of individual songs affects playlist experience. To study the relationship between musical sophistication and music listening behavior, participants' logged into our app through the Spotify API, which allowed us to retrieve their music listening behavior. In addition, participants completed a survey with items from the Goldmiths Musical Sophistication Index (Gold-MSI;~\cite{mullensiefen2014musicality}). The Gold-MSI measures musical sophistication on five subscales: active engagement, emotions, singing abilities, perceptual abilities, and musical training. However, in this preliminary work, we only asked participants to respond on two of these subscales as we believe that they are the most prominent ones reflected in online music listening behaviors: 
\begin{enumerate}
	\item Active engagement (e.g., how much time and money one spends on music; measured by 9-items). 
	\item Emotions (e.g., active behaviors related to emotional responses to music; measured by 6-items).
\end{enumerate}

A total of 61 participants were recruited in December 2017 through a participant pool managed by the Human-Technology Interaction group at Eindhoven University of Technology: 28 male, 33 female (mean age: 23.92 years, SD: 4.57 years). Using Spotify's API we retrieved the participants' top tracks, which resulted in a dataset of 21,080 tracks. For each track we retrieved the audio features through Spotify's API: valence (0-1: negative-positive emotions), liveness, instrumentalness, energy (0-1: calm-energetic), danceability, tempo (BPM), time signature, loudness (dB), track popularity  and artist popularity. \footnote{Popularity measures are not explained in detail by Spotify, but range from 0 to 100.} For each feature we calculated the standard deviation, mean, median, min and max values.

\section{Results}
We used a learner-based feature selection to select the best features (track properties) to create a model to predict participants' emotions and active engagement scores~\cite{mullensiefen2014musicality} from their music listening behaviors. A ZeroR classifier was used to create a baseline predictive model. Two different classifiers were used and compared against the baseline model: random forest and radial basis function network (RBF network). Each classifier was applied to the selected features (see Table~\ref{features}).

Our predictive models were trained with the aforementioned classifiers in Weka \citep{Hall:2009:WDM:1656274.1656278} with a 10-fold cross-validation with 10 iterations. For each classifier used, we report the root-mean-square error (RMSE) in Table~\ref{classifier} to indicate the root mean square difference between predicted and observed values. The RMSE of each music sophistication trait relates to a [1,7] score scale.

\begin{table}[h]
\centering
\begin{tabular}{ll}
\textbf{Emotions}              & \textbf{Active Engagement} \\ \hline
Valence std.dev          & Track popularity std.dev        \\
Tempo std.dev            & Valence mean             \\
Time signature mean & Valence median           \\
Time signature std.dev  & Valence max              \\
Time signature min  & Tempo std.dev                \\
Liveness mean        & Time signature std.dev      \\
Liveness std.dev         & Time signature min      \\
Liveness median      & Loudness median          \\
Instrumentalness std.dev & Energy max               \\
Energy std.dev           & Danceability mean        \\
Energy min           & Danceability median      \\
Danceability mean    &                           \\
Danceability std.dev     &                           \\
Danceability median  &                           \\
Danceability min     &                          
\end{tabular}
\caption{Selected features for the predictive models.}
\label{features}
\end{table}

We first trained a random forest classifier. Random forests have shown to have a reasonable performance when the features consist of high amounts of noise~\cite{humston2010quantitative}. As the random forest classifier failed to outperform the baseline in the emotions dimension, we used the RBF network classifier. The RBF network is a neural network that has shown to work well on smaller datasets~\cite{khot2012evaluation}.

\begin{table}[h]
\centering
\begin{tabular}{lccc}
                          & \textbf{ZeroR} & \textbf{Random Forest} & \textbf{RBF network}    \\ \hline
Emotions      & 0.97  & 0.99 & \textbf{0.95} \\
Active Eng. & 0.97  & \textbf{0.93} & \textbf{0.93}
\end{tabular}
\caption{RMSE scores ($r$~$\in$~[1,7]) of predicting emotions and active musical engagement from listening behavior. Boldfaced numbers indicate an out performance of the baseline.}
\label{classifier}
\end{table}

\section{Conclusion, Limitations \& Outlook}
In this preliminary work we explored the prediction of musical sophistication subscales (i.e., emotions and active engagement) from music listening behavior. Our results show that music listening behavior can be used to infer the musical sophistication of users. We used a random forest classifier and an RBF network classifier to create the predictive models. Although both classifiers were able to outperform the baseline model on active engagement prediction, only the RBF network was able to also outperform the baseline on predicting the emotions subscale.

Although we were able to predict participants' scores on two subscales of Gold-MSI from music listening behavior, performance can likely be improved more. To do this, we plan to extend the analysis in several ways. We aim to expand our dataset by increasing the number of participants in our dataset and the number of measurements per participant. This will allow for a more in-depth investigation of the relationship between behavioral features and the Gold-MSI scores. Furthermore, we plan to explore the prediction of other subscales of the Gold-MSI as well as exploring the predictive value of other music listening behaviors that are available through the Spotify API (e.g., user's playlists and social networks).

\section{Acknowledgements}
We would like to thank Eelco Wiechert for creating the application.

\bibliographystyle{ACM-Reference-Format}
\bibliography{acmart}

\end{document}